\documentclass[12pt]{iopart}

\pdfoutput=1

\usepackage{iopams}  
\usepackage[pdftex]{graphicx}
\usepackage{hyperref}
\usepackage{amstext}
\usepackage{amssymb}
\usepackage[usenames]{color}

\begin{document}

\title[Longitudinal momentum spread after tunnel ionization]{Comparison of different approaches to the longitudinal momentum spread after tunnel ionization}

\author{C. Hofmann, A. S. Landsman, C. Cirelli, A. N. Pfeiffer, U. Keller}

\address{Physics Department, ETH Zurich, CH-8093 Zurich, Switzerland}
\ead{chofmann@phys.ethz.ch}

\begin{abstract}
We introduce a method to investigate the longitudinal momentum spread resulting from strong-field tunnel ionization of Helium which, unlike other methods, is valid for all ellipticities of laser pulse. Semiclassical models consisting of tunnel ionization followed by classical propagation in the combined ion and laser field reproduce the experimental results if an initial longitudinal spread at the tunnel exit is included.  The values for this spread are found to be of the order of twice the transverse momentum spread.
\end{abstract}

\pacs{33.20.Xx, 31.15.xg, 32.80.Rm, 33.60.+q} 
\maketitle

\section{Introduction}

Applying a strong electromagnetic field bends the binding potential of an atom or a molecule, allowing the electron to tunnel out.  In attosecond science \cite{AttosecondPhysics}, this tunnel ionization is commonly assumed to be the starting point of many important phenomena, including High Harmonic Generation (HHG).  Consistent semiclassical models are required to interpret and explain results from ultrafast laser experiments \cite{TunnelDelay,Attoclock} as well as help design new experiments.

An assumption that comes from the tunnelling limit of the Strong Field Approximation (SFA), is that at the tunnel exit, an electron does not have any momentum parallel to the electric field (see for example \cite{PRL:233001}). This is contrary to the transverse momentum spread, for which the ADK theory \cite{ADK,Delone1991,Popov2004}
predicts a Gaussian distribution with a standard deviation given by
\begin{equation}
\sigma_{\bot}= \sqrt{\frac{\omega}{2\gamma(t_0)}},  \label{eq:vtransProb}
\end{equation}
where $\omega$ is the laser frequency, $\gamma(t_0) = \omega \sqrt{2I_{\mathrm{p}}}/F(t_0)$ the Keldysh parameter \cite{Keldysh1965} at ionization time $t_0$, $F(t_0)$ is the field strength at ionization time and $I_{\mathrm{p}}$ is the ionization potential of the atom.

In \cite{Wavepacket} experimental evidence was presented for the existence of the momentum distribution parallel to the electric field at the tunnel exit.  We follow up on this result by presenting a detailed analysis of semiclassical simulations and their comparison to experimental results using a newly developed  method of elliptical integration, which is robust at all ellipticities $\epsilon$.  Other methods, such as radial integration, work well at high ellipticities, but break down for low $\epsilon$, mixing transverse and longitudinal components of the momentum spread.  On the other hand, projection onto the main axis of polarization works well at low $\epsilon$, but breaks down at higher $\epsilon$.

The elliptical integration method confirms in agreement with \cite{Wavepacket}, that an initial longitudinal momentum spread at the tunnel exit fits in well with our experimental results.  This initial spread was found to be larger than the transverse spread given by \eref{eq:vtransProb}.

\section{Semiclassical model}
After the quantum mechanical tunnel ionization, the interaction of the freed electron with the laser field is considered classically. 
The electric field of the pulse can be written as 
\begin{equation}
\bi{F}(t) = \frac{F_0}{\sqrt{1+\epsilon^2}} \left( \cos(\omega t + \varphi_{\mathrm{CEO}}) \hat\bi{{x}} +
\epsilon\sin(\omega t + \varphi_{\mathrm{CEO}})\hat\bi{{y}}  \right)f(t), \label{eq:pulse}
\end{equation}
where $f(t)$ is the pulse envelope, the $x$-axis is the major axis of the polarization ellipse and the $y$-axis is the minor axis of the polarization ellipse. 
The momentum spread of the electron wave packet at the tunnel exit transverse to the electric field follows from the ionization rate calculation. Formally, the kinetic energy from the transverse momentum adds to the ionization potential \cite{Anatomy}. In a quasistatic picture, the ionization probability at the tunnel exit can thus be expressed as follows \cite{Anatomy}, 
\begin{equation} 
P(v_x=0,v_y,v_z) \approx \exp \left( \frac{-2 (2 I_{\mathrm{p}})^{3/2}}{3F} \right) \exp \left(- \frac{v_y^2 + v_z^2}{2\sigma_{\bot}^2} \right), \label{eq:IonProbabTransverse}
\end{equation}
where the longitudinal momentum at the tunnel exit parallel to the electric field, $v_x$, is typically assumed to be zero. The transverse momentum spread is given by \cite{Delone1991,Popov2004}
\begin{equation}
\sigma_{\bot}^2 = \frac{F}{2(2I_{\mathrm{p}})^{1/2}} = \frac{\omega}{2\gamma},  \label{eq:ADKtransverse}
\end{equation}
and the exponential was expanded to give two separate terms for the ionization potential and the transverse momentum distribution.
This transverse spread is valid directly at the tunnel exit point as well as at the detector after propagation in the laser field, since the field exerts virtually no net force in the transverse direction and the influence of the ion Coulomb field is neglected after ionization in SFA \cite{Anatomy}.

\begin{figure}[htb]
\centering
(a) \hspace{0pt}\raisebox{-0.1\height}{\includegraphics[width=0.35\textwidth]{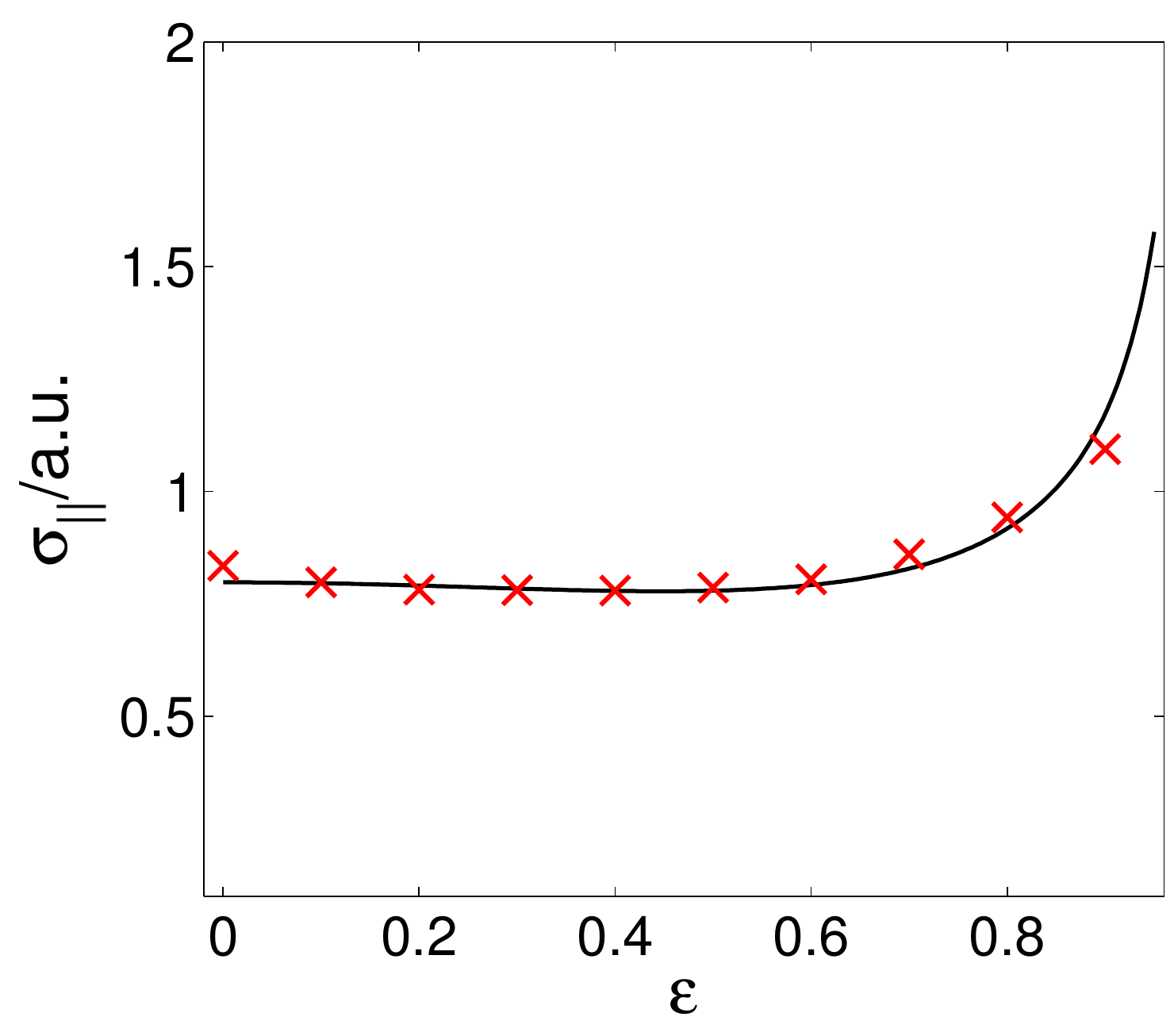}} \hspace{20pt}
(b) \hspace{0pt}\raisebox{-0.1\height}{\includegraphics[width=0.35\textwidth]{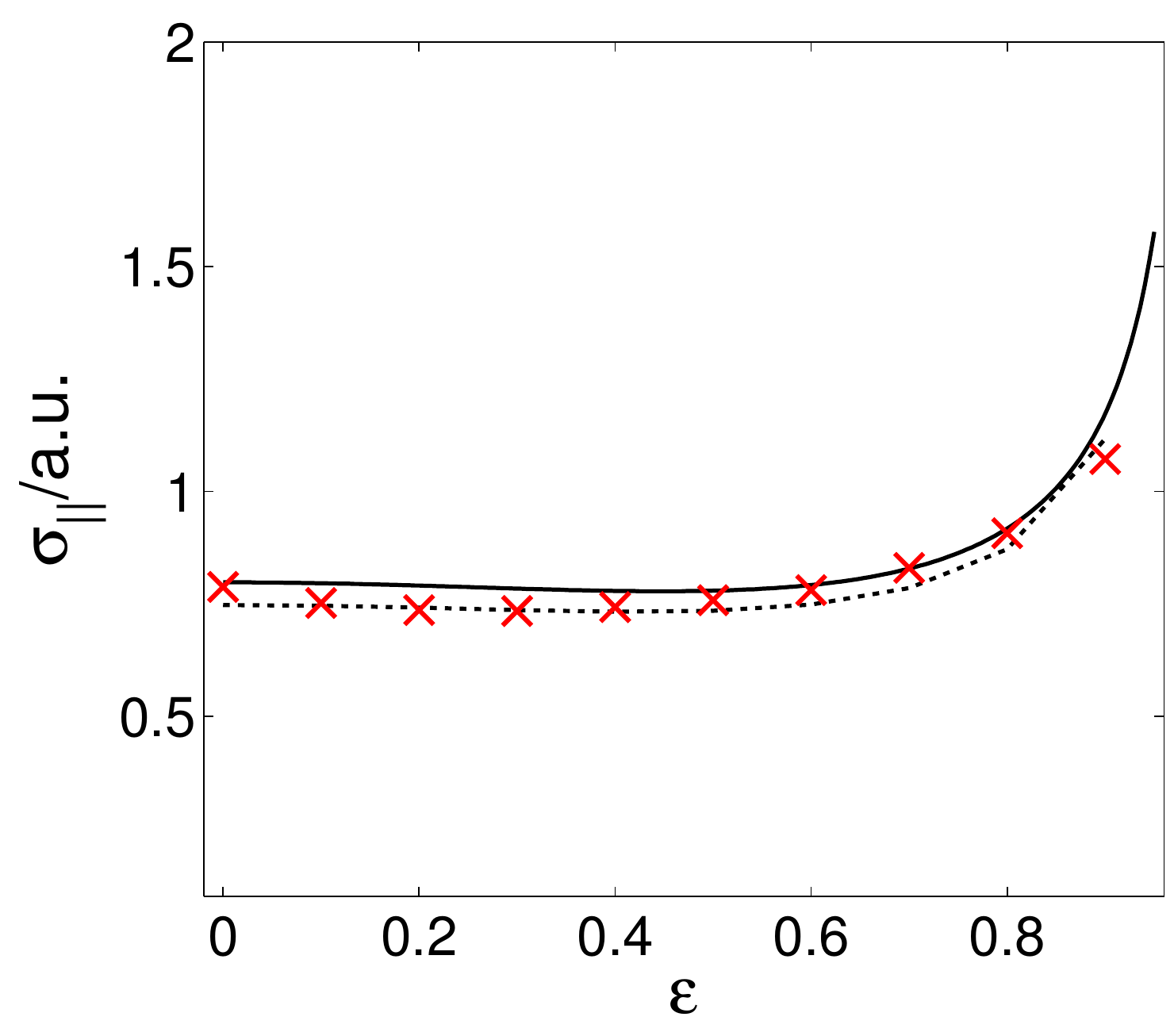}}
\caption{\textbf{Final longitudinal momentum spread.} The analytical formula \eref{eq:sigmaLTheo} is drawn as a black solid line in both plots. (a) Red $\times$ show values obtained from Monte Carlo simulations where the ionization events were restricted to the optical cycle at the peak of the pulse. (b) The black dotted curve is the averaged analytical formula \eref{eq:sigmaLTheoAverage}. Red $\times$ show values obtained from Monte Carlo simulations where ionizations are spread over the pulse.}
\label{fig:figure01}
\end{figure}

On the other hand, there is no clear way of calculating the longitudinal momentum spread at the exit point from tunnelling models. But, a theoretical estimate for the longitudinal momentum spread at the detector \cite{JETP:EMspectra}
\begin{equation}
\sigma_{||}^{\mathrm{final}}=\sqrt{\frac{3\omega}{2\gamma^3(1-\epsilon^2)}}, \label{eq:sigmaLTheo}
\end{equation} 
can be calculated, assuming zero initial longitudinal momentum. This spread is due to the different phases of the laser field at ionization time. Electrons ionized before or after a peak of the field feel a net force due to the field, which leads to a spread of longitudinal momentum acquired during propagation in the laser field. This is verified in figure \ref{fig:figure01}, where \eref{eq:sigmaLTheo} is compared to the results from a simulation (find details in section \ref{CTMC}) showing very good agreement for $\epsilon$ not at extreme values.

The middle optical cycle has the maximum peak field strength, which decreases for optical cycles outside the pulse centre. Averaging the variances from \eref{eq:sigmaLTheo} with weights given by the ionization probability at the peak of each optical cycle $P(p)$ from \eref{eq:IonProbab} yields a slightly smaller overall longitudinal spread
\begin{equation}
\overline{\sigma_{||}^{\mathrm{final}}}= \left(  \frac{\sum P(p)(\sigma_{||}^{\mathrm{final}}(p))^2  }{\sum P(p)}  \right)^{1/2} \label{eq:sigmaLTheoAverage}.
\end{equation}

\section{Semiclassical Simulation} \label{CTMC}
All simulations use atomic units. A Classical Trajectory Monte Carlo (CTMC) simulation was performed to investigate how the standard assumption of $ \sigma_{||}^{\mathrm{initial}}=0$ for the longitudinal momentum has to be adapted to fit the experiment. For varying initial momentum spread at the exit point, the final momentum distribution was calculated.

\begin{figure}[htb]
\centering
\includegraphics[width=0.5\textwidth]{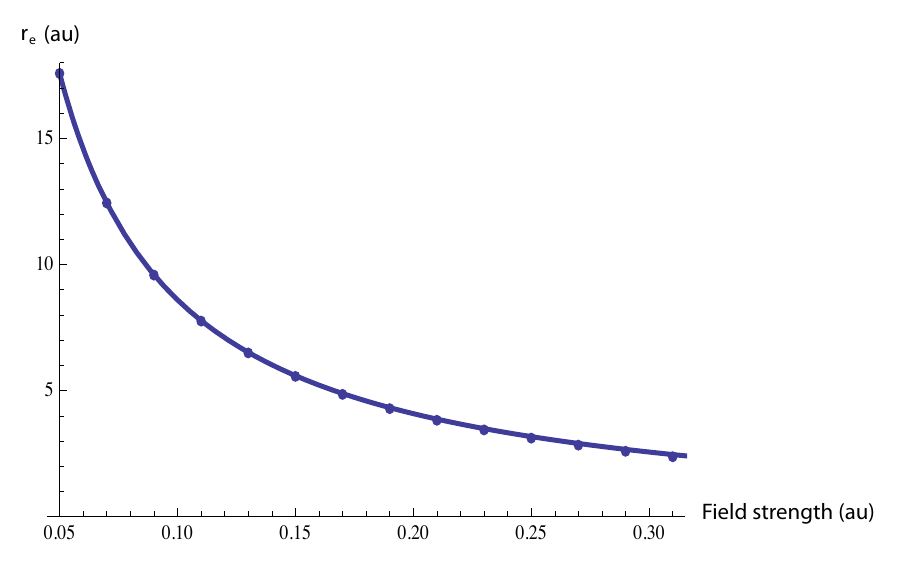}
\caption{\textbf{Tunnel exit radius.} The exit radius calculated from the quadratic approximation given by \eref{eq:quadraticExit} (solid line) is compared to the full cubic solution in parabolic coordinates (dots).}
\label{fig:figure02}
\end{figure}

\begin{figure}[htb]
\centering
\includegraphics[width=0.35\textwidth]{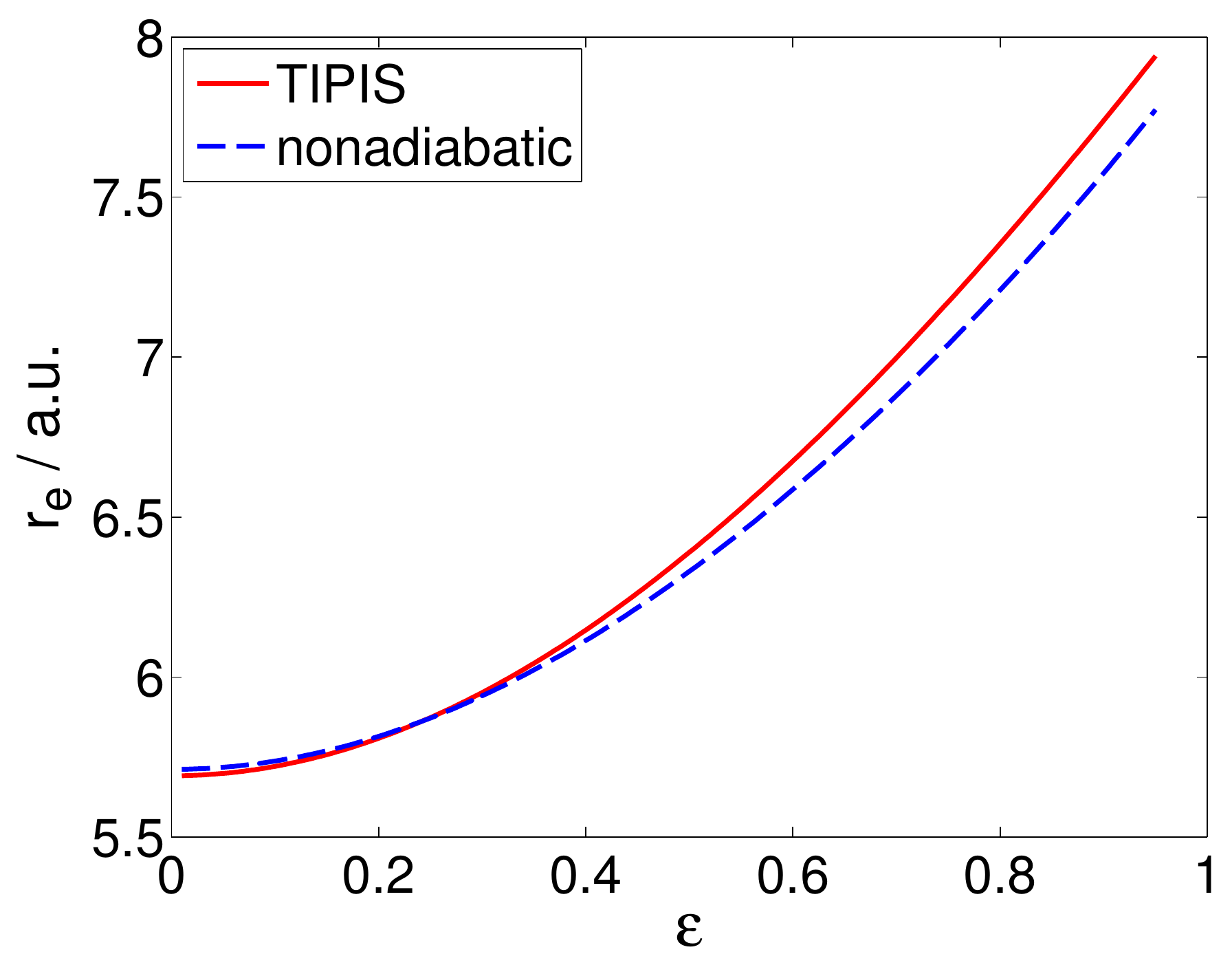}
\caption{\textbf{Non-adiabatic tunnel exit radius.} The exit radius as calculated by the TIPIS model \eref{eq:quadraticExit} (red solid line) is compared to the values found using non-adiabatic theory \cite{PPT2} (blue dashed line) over the complete range of ellipticity. }
\label{fig:figure}
\end{figure}

The simulations follows the TIPIS model (Tunnel Ionization in Parabolic coordinates with Induced dipole and Stark shift) \cite{Attoclock}. 
The Stark shifted ionization potential is given by
\begin{equation}
I_{\mathrm{p}} \equiv I_{\mathrm{p}}(F(t_0)) = I_{\mathrm{p},0} + \frac{1}{2} \left(\alpha_{\mathrm{N}} - \alpha_{\mathrm{I}}\right)F(t_0)^2,
\end{equation}
where $\alpha_{\mathrm{N}}$ and $\alpha_{\mathrm{I}}$ denote the polarizability of the atom and the ion respectively. The equation of motion, including the induced dipole in the ion due to the laser field
\begin{equation}
\ddot{\bi{r}}(t) = - \bi{F}(t) - \nabla \left[V(\bi{r},t)\right], \qquad V(\bi{r},t) = \frac{-1}{r}  - \alpha_{\mathrm{I}}\frac{\bi{F}\cdot\bi{r}}{r^3} \label{eq:fullEquation}
\end{equation}
is solved numerically to compute the trajectory of electrons post ionization. 
The simulation uses a $\cos^2$ laser pulse centred about $t=0$ with a complete pulse duration of 90 fs, a wavelength of 788 nm and $F_0 = 0.15 \mathrm{\, a.u.}$. These parameters are chosen to match the experiment in \cite{Wavepacket}. Ensembles of $3\times 10^5$ events are created for each ellipticity and longitudinal momentum spread combination. All electrons have a random exit time, with probability weighted by \cite{Keldysh1965} 
\begin{equation}
P(t_0) \propto \exp\left(-\frac{2(2I_p)^{3/2}}{3 F(t_0)} \right)  \label{eq:IonProbab},
\end{equation}
and an instantaneous tunnelling time is assumed. For the initial condition the TIPIS model \cite{Attoclock} is used, which calculates the tunnel exit point  by solving the cubic equation that gives the potential in the parabolic coordinate $\eta$, where $\eta$ corresponds to the direction of tunnelling.  For $\eta \gg 1$ (corresponding to exit radius $r_e > 5 \mathrm{\,a.u.}$, a condition satisfied by present day strong field ionization experiments), the cubic potential in the direction of tunnelling is well-approximated by a quadratic, since the terms $\propto 1/\eta^2$ in the potential described in \cite{Attoclock} can be neglected. This quadratic approximation yields
\begin{equation}
r_e = \frac{\eta_e}{2} = \frac{I_p + \sqrt{I_p^2 - 4\beta_2F(t_0)}}{2 F(t_0)} \label{eq:quadraticExit}
\end{equation}
for the tunnel exit radius, while $\beta_2$ is given by
\begin{equation}
\beta_2 = 1 - \frac{\sqrt{2 I_p}}{2}.
\end{equation}
Figure \ref{fig:figure02} shows that the exit radius given by the full solution of the potential in parabolic coordinates and the quadratic approximation show excellent agreement over a wide range of electric field strengths.
Additionally, the tunnel exit coordinates given by \eref{eq:quadraticExit} agree within 2.2\% with non-adiabatic theory \cite{PPT2} for our experimental parameters (see section \ref{experiment}). Figure \ref{fig:figure} shows a comparison of the tunnel exit points predicted by the TIPIS model with non-adiabatic, gamma-dependent values given in \cite{PPT2}.

To approximate the ion potential \eref{eq:fullEquation}, a soft core potential \cite{PRL:233001} with $ a^2 = 0.1 \mathrm{\, a.u.}$ is implemented.

At any instant, the direction of the laser field defines a coordinate system with basis 
$\{b_{||}, b_{\bot,\mathrm{ip}}, b_{\bot,\mathrm{op}}\}$, parallel to the field, orthogonal to the field but in the plane of polarization, and orthogonal to the plane of polarization. For both $v_{\bot,\mathrm{ip}}^{\mathrm{initial}}$ and $v_{\bot,\mathrm{op}}^{\mathrm{initial}}$, Gaussian distributed values are generated independently using the standard deviation given by \eref{eq:ADKtransverse}, resulting in the transverse momentum distribution
\begin{equation} 
P(v_{\bot,\mathrm{ip}},v_{\bot,\mathrm{op}}) \propto\exp \left(- \frac{(v_{\bot,\mathrm{ip}}^{\mathrm{initial}})^2}{2\sigma_{\bot}^2} \right)\exp \left(- \frac{(v_{\bot,\mathrm{op}}^{\mathrm{initial}})^2}{2\sigma_{\bot}^2} \right). \label{eq:TransMomComp}
\end{equation}

\begin{figure}[htb]
\centering
\includegraphics[width=\textwidth]{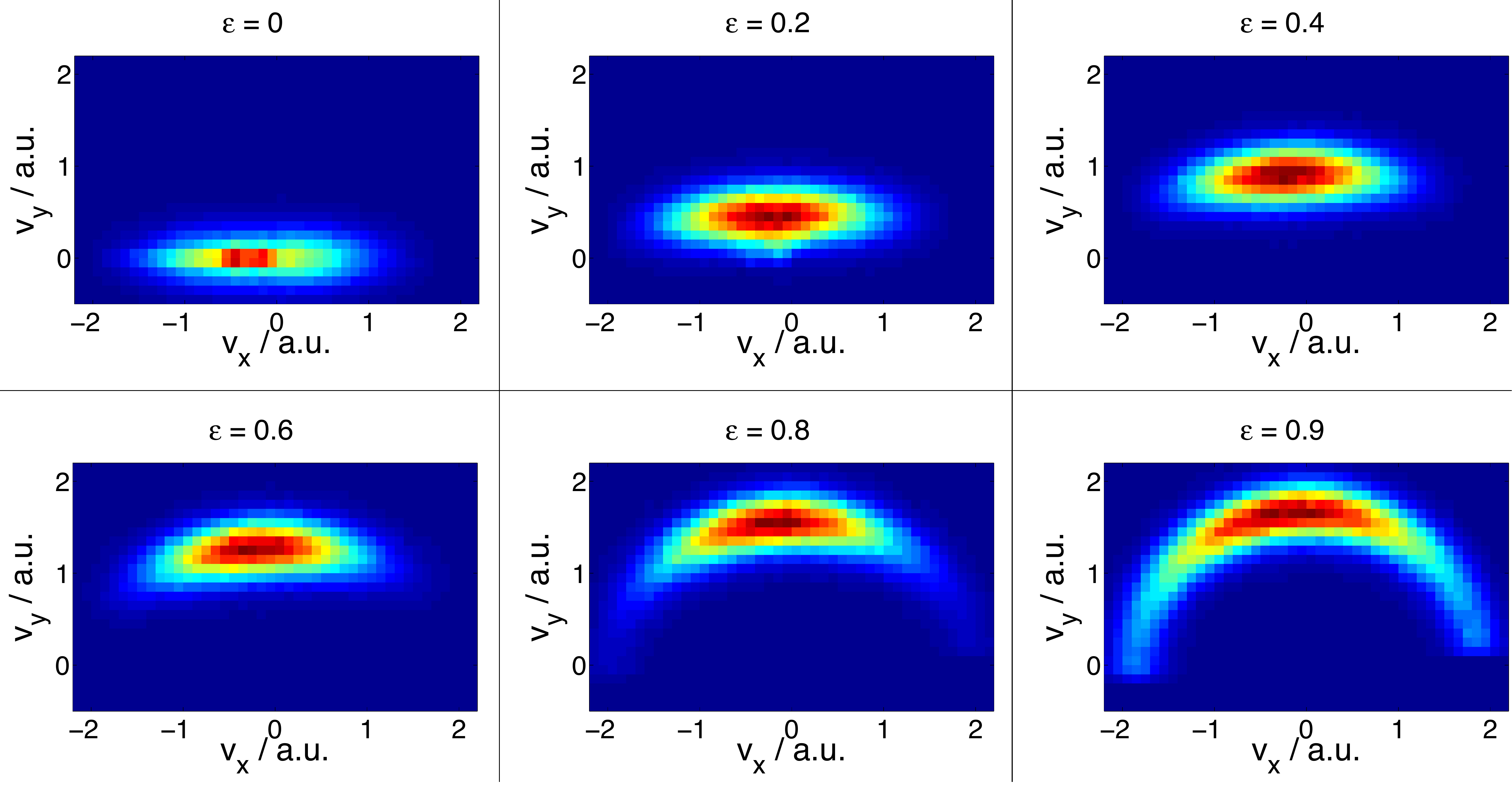}
\caption{\textbf{Electron momentum distribution scan over ellipticity.} For six different ellipticities $\epsilon$, the momentum distribution of ionized electrons projected onto the plane of polarization is shown, where $x$ is the major axis of polarization and $y$ the minor axis. The slight anticlockwise tilt of the centres is due to the Coulomb interaction with the parent ion \cite{Attoclock}.}
\label{fig:SimScanEllipticity}
\end{figure}

Figure \ref{fig:SimScanEllipticity} shows momentum distributions calculated from the simulation. As expected, with increasing ellipticity the two centres of distribution move to higher transverse momentum and the distribution approaches a doughnut shape. For the calculation of the distribution, Rydberg electrons \cite{PRL:233001,Landsman2013} and electrons which came closer than $5 \mathrm{\,a.u.}$ during the pulse are discarded, because at these length scales the classical model fails in reconstructing quantum mechanical interactions between the electron and the ion.

\section{Analysis of final longitudinal momentum distribution} \label{comparison}
Here we describe three different methods for analysing the final longitudinal momentum spread:  1) radial integration, 2) projection onto the major axis of polarization and 3) elliptical segment integration.  The intent of all three integration methods is to integrate over the transverse momenta, so that the resulting one-dimensional distributions are functions of the longitudinal spread only.
The first two methods break down at low and high ellipticities, respectively, while elliptical integration holds over the entire ellipticity range $|\epsilon| < 1$. At fully circular polarization $\epsilon = \pm 1$, the momentum distribution is isotropic and no information about the longitudinal spread can be extracted. This is mirrored in the fact that the analytical formula \eref{eq:sigmaLTheo} diverges as $|\epsilon|$ approaches 1.

\subsection{Radial integration}
Previously, the two dimensional momentum distribution in the plane of polarization was integrated radially, and the resulting angular distribution was compared \cite{Wavepacket}. The radial integration is easy to implement and effective for high ellipticity, where it focuses on the spread parallel to the field, integrating over the transverse spread. With decreasing ellipticity, however, the radial integration mixes transverse and longitudinal components more and more as the centre of distribution moves closer to the origin (see figure \ref{fig:SimScanEllipticity}). This causes problems with reliability in the fitting process, and the angular distribution starts to develop a spurious double peak structure, where there is only one peak in the two-dimensional distribution (see for example figure \ref{fig:figCompP}).  Therefore, the radial integration method is only reliable at higher ellipticities, increasing in accuracy as $\epsilon$ increases.

\subsection{Projection onto the major axis of polarization} 
As an alternative, the momentum distribution can be integrated over both transverse momentum distributions (out of plane of polarization $z$ and minor polarization axis $y$), thus creating a momentum distribution projected onto the major axis of polarization, $x$. This method works well for small ellipticities, where the momentum distribution is only slightly curved.  However, with increasing $\epsilon$, the accuracy of this projection breaks down, as it begins to integrate over the longitudinal spread, particularly at higher absolute values of $v_x$ (see figure \ref{fig:SimScanEllipticity}).

\subsection{Elliptical segment integration}
\begin{figure}[htb]
\centering
\includegraphics[width=0.35\textwidth]{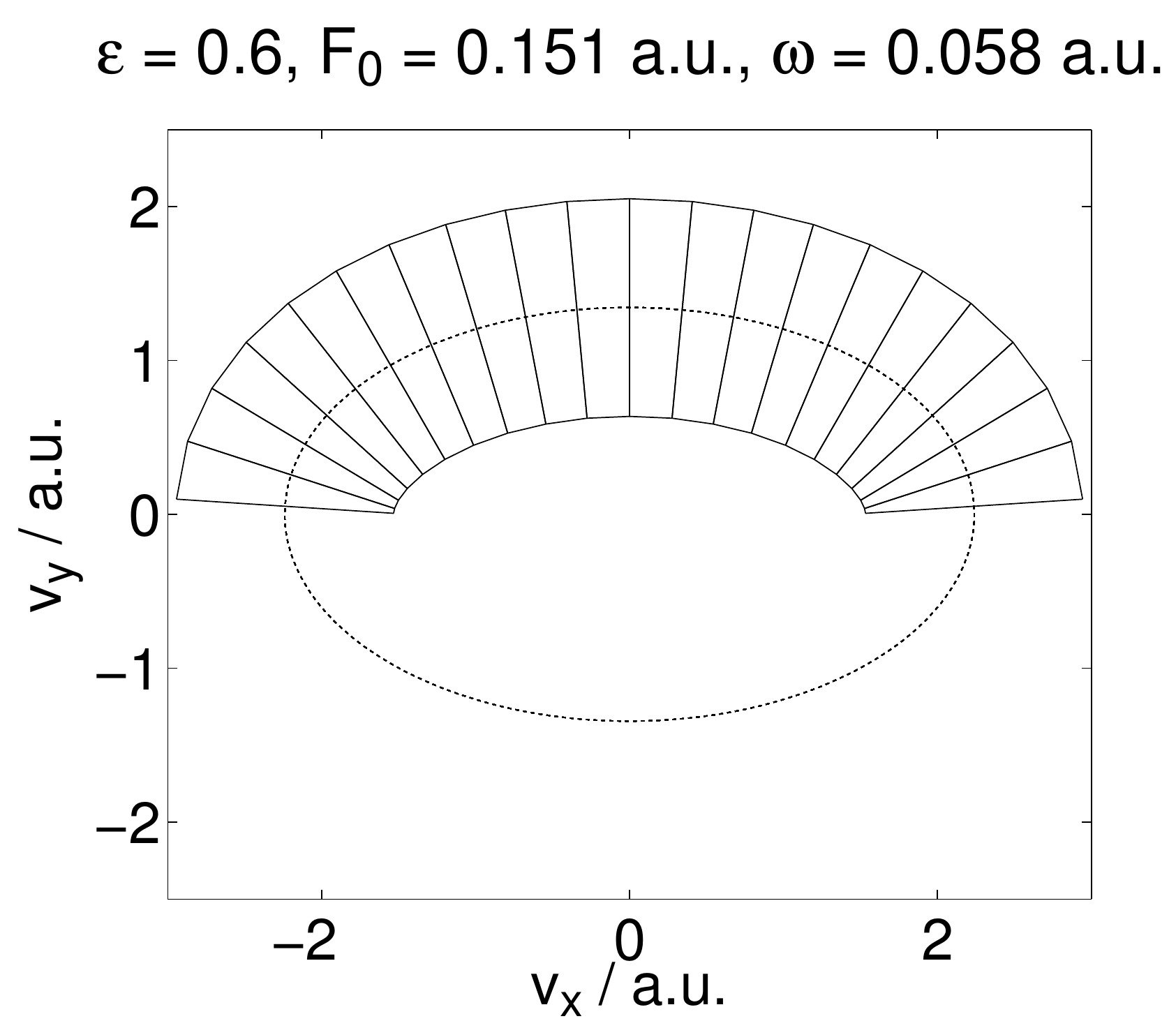}
\caption{\textbf{Elliptical segments.} The final momentum distributions lie around the ellipse shown as a dotted line. The plotted segments are orthogonal to the elliptic curve, denoting the segments for elliptical binning of the data to study the longitudinal distribution parallel to the ellipse.}
\label{fig:figure_04}
\end{figure}

It is desirable to have a single technique for extracting the longitudinal spread applicable over the whole range of ellipticities. It must take into account that the momentum distribution is centred on an ellipse, with eccentricity given by $\epsilon$. To analyse the longitudinal momentum distribution, an integration over lines perpendicular to this ellipse has to be performed. The momentum distribution as a function of angle is then given by, 
\begin{eqnarray}
P(\phi) = \sqrt{1+\epsilon^2\tan^2\phi}  \int_{y_-}^{y_+} f(x_0 - \epsilon\tan\phi(y-y_0),y) \,\rmd y, \\ 
	y_{\pm} = y_0 \pm \frac{1}{2}(1+\epsilon^2 \tan^2\phi)^{-1/2} \nonumber \\
	x_0(\phi) = \frac{-F_0}{\omega\sqrt{1+\epsilon^2}}\sin\phi, \qquad y_0 = \frac{F_0\epsilon}{\omega\sqrt{1+\epsilon^2}}\cos\phi, \nonumber
\end{eqnarray}
where $\phi$ is counted from the $y$ axis counter clockwise and $f(x,y)\equiv f(v_x,v_y)$ is the momentum probability density in the plane of polarization. In the numerical approximation, this corresponds to binning all events into the ellipse segments pictured in figure \ref{fig:figure_04} and weighting the results with the inverse area of the corresponding segment. 

The longitudinal momentum spread at the detector can easily be extracted from this technique. Fitting a Gaussian to the resulting momentum distribution yields an angle of highest probability $\phi_{\mathrm{m}}$ and the standard deviation in angle $\sigma_{\phi}$. From this, the longitudinal momentum spread is given by
\begin{equation}
\sigma_{||}^{\mathrm{final}} = \frac{1}{2} \frac{F_0}{\omega\sqrt{1+\epsilon^2}} \int_{\phi_{\mathrm{m}} - \sigma_{\phi}}^{\phi_{\mathrm{m}} + \sigma_{\phi}} \sqrt{\cos^2\phi + \epsilon^2\sin^2\phi} \;\rmd \phi.
\end{equation}

\section{Experimental Setup}  \label{experiment}
A COLd Target Recoil Ion Momentum Spectroscopy (COLTRIMS) \cite{Doerner2000} setup measures the ion momentum of Helium ions, which is the negative of the electron momentum due to momentum conservation. In the COLTRIMS setup, the fragments created in the interaction region are guided towards time and position sensitive detectors by constant electric and magnetic fields. The raw data, consisting of the time-of-flight and positions of impact of ions and electrons, is used to calibrate the electric and magnetic fields. Then, the electron and ion momenta are calculated by solving the equations of motion.

The momentum resolution is 0.1 a.u. in time-of-flight direction and estimated to 0.9 a.u. in gas jet direction, mainly determined by thermal spread of the gas jet. The momentum resolution is better in the lower half of the detector plate, therefore we consider only ions that impact in the lower part of the detector plate and mirror the distribution onto the upper half. This makes the spectra perfectly point-symmetric. 

A Titanium:Sapphire based laser system produces short intense laser pulses with a pulse duration of 33 fs (FWHM) at a central wavelength of 788 nm. The pulse field can be approximated by \eref{eq:pulse}.
The carrier-envelope-offset (CEO) phase $\varphi_{\mathrm{CEO}}$ \cite{Telle1999} is not stabilized. The peak intensity $I = F_0^2$ is estimated to be $0.8\times 10^{15} \mathrm{\,W/cm^2}$ by matching a Monte Carlo simulation described in \cite{Wavepacket} to the data regarding the momentum distribution along the y-axis \cite{Alnaser2004} (the atomic unit of intensity is $3.509 \times 10^{16} \mathrm{\,W/cm^2}$). The Keldysh parameter depends on the ellipticity and ranges between $\gamma = 0.51$ for $\epsilon = 0$ and $\gamma = 0.73$ for $\epsilon = 1$.

The polarimetry of the experiment has been described elsewhere \cite{Attoclock}. Using a broadband quarter-wave plate, ellipticities up to $\epsilon \pm 0.93$ can be achieved.
While recording the COLTRIMS data, the quarter-wave plate is rotated continuously by a motorized rotary stage and the angle is read out and tagged to the dataset for each laser pulse. The angular orientation of the polarization ellipse is calculated for each measured ion, and in the presentation of the data the $x$-axis designates always the major polarization axis rather than an axis fixed in laboratory space. The calculation of the ellipticity allows generating ellipticity-resolved spectra with a high resolution. 

\section{Measurements}

\begin{figure}[htb]
\centering
\includegraphics[width=\textwidth]{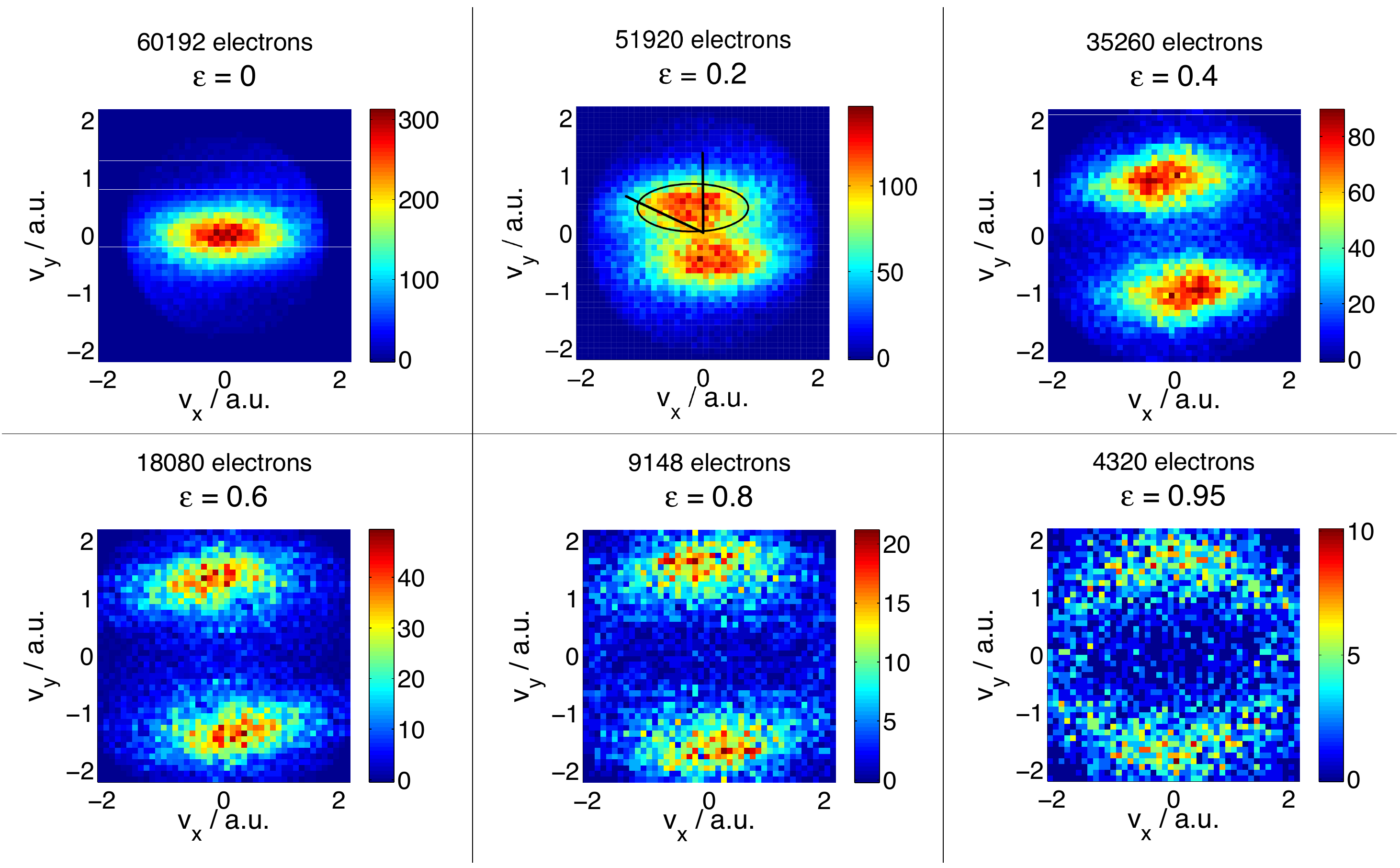}
\caption{\textbf{Ion momentum distribution scan over ellipticity.} For six different ellipticities $\epsilon$, the momentum distribution of ionized electrons as recorded by the COLTRIMS setup is shown, where $x$ is the major axis of polarization and $y$ the minor axis. With increasing ellipticity, the two centres of distribution move to higher transverse momentum and the distribution approaches a doughnut shape. The slight anticlockwise tilt of the centres is due to the Coulomb interaction with the parent ion \cite{Attoclock}.}
\label{fig:MomScanEllipticity}
\end{figure}

Figure \ref{fig:MomScanEllipticity} shows the momentum distribution projected onto the plane of polarization in the case of anticlockwise rotating field. For each ellipticity, recorded events in an interval of $\pm 0.025$ around the indicated ellipticity were integrated.  

For $\epsilon>0$, the two main distributions split apart and lengthen with increasing ellipticity, to form a near circular distribution at $\epsilon \approx 1$. 
Results from the clockwise rotating field should be mirror symmetric with respect to the $y$-axis, compared to counter-clockwise field. They were recorded as well and included in quantitative comparisons to simulation data (see \ref{comparison}) to reduce systematic errors.

\section{Longitudinal momentum spread}
\begin{figure}[h!]
\centering
\includegraphics[width=\textwidth]{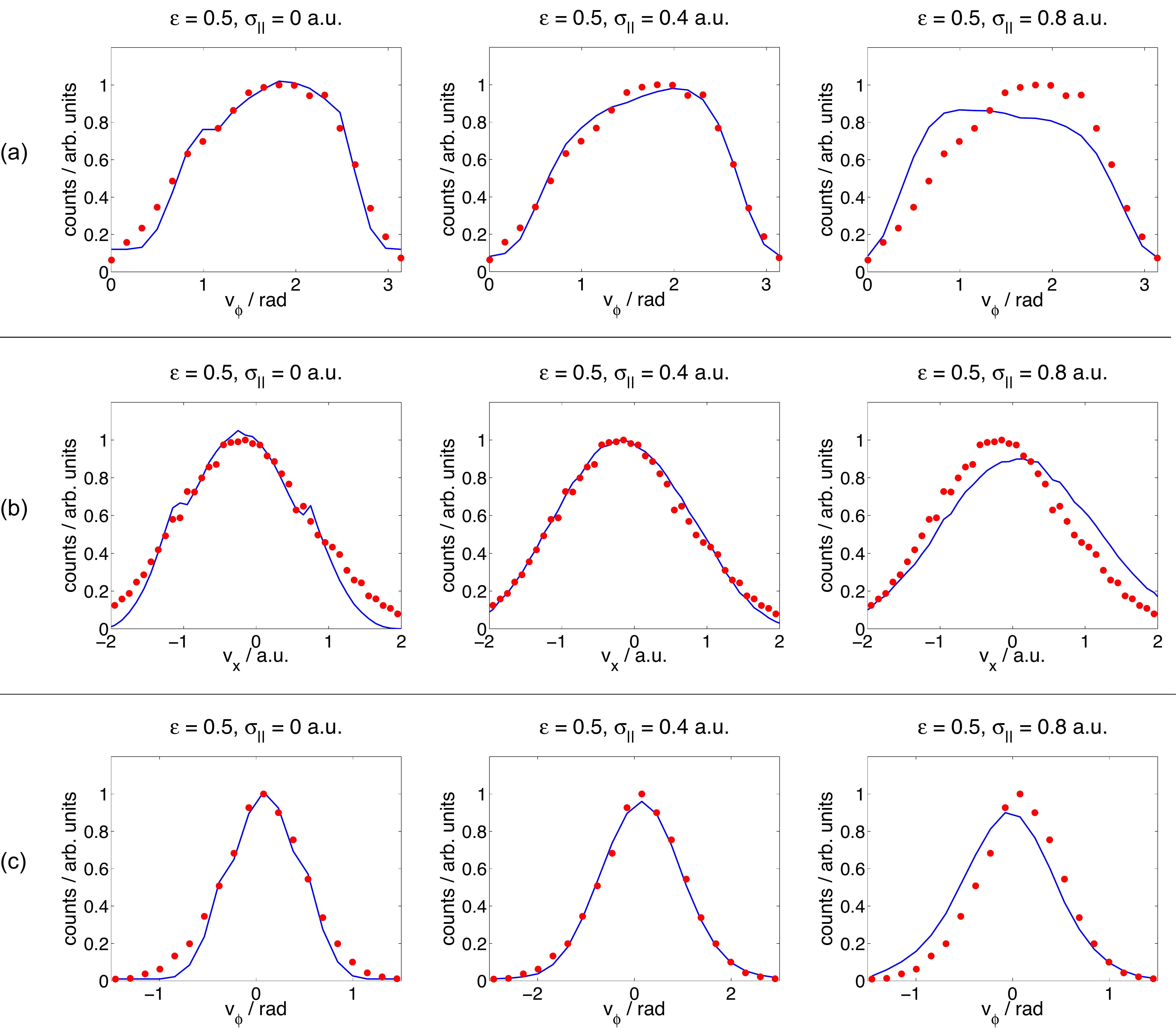}
\caption{\textbf{Momentum distribution comparison for the top half.} Red dots are the experimental data for ellipticity $\epsilon = 0.5$, blue solid lines show the results from the simulation where initial longitudinal momentum spreads of 0 a.u., 0.4 a.u. and 0.8 a.u. were used. (a) Comparison using radial integration, $\phi$ counted anticlockwise from the $x$-axis. (b) Comparison using $x$ projection. (c) Comparison using the elliptical segments technique, $\phi$ counted anticlockwise from the $y$-axis.}
\label{fig:e5_figComp} 
\end{figure}

In order to find the best fitting initial longitudinal momentum spread from the simulation, quantitative comparisons between experimental data and simulation results were calculated. Both the experimental momentum distribution as well as the resulting distributions from the simulation were analysed using all of the above discussed techniques. For each ellipticity, a set of simulations with varying initial longitudinal momentum spread, step size 0.05 a.u., was calculated. 
Figure \ref{fig:e5_figComp} shows an example of a comparison between the simulation and the experimental data for the case of ellipticity 0.5 as calculated using radial integration (a), $x$ projection (b) and elliptical integration (c).

\begin{figure}[h]
\centering
\includegraphics[width=0.4\textwidth]{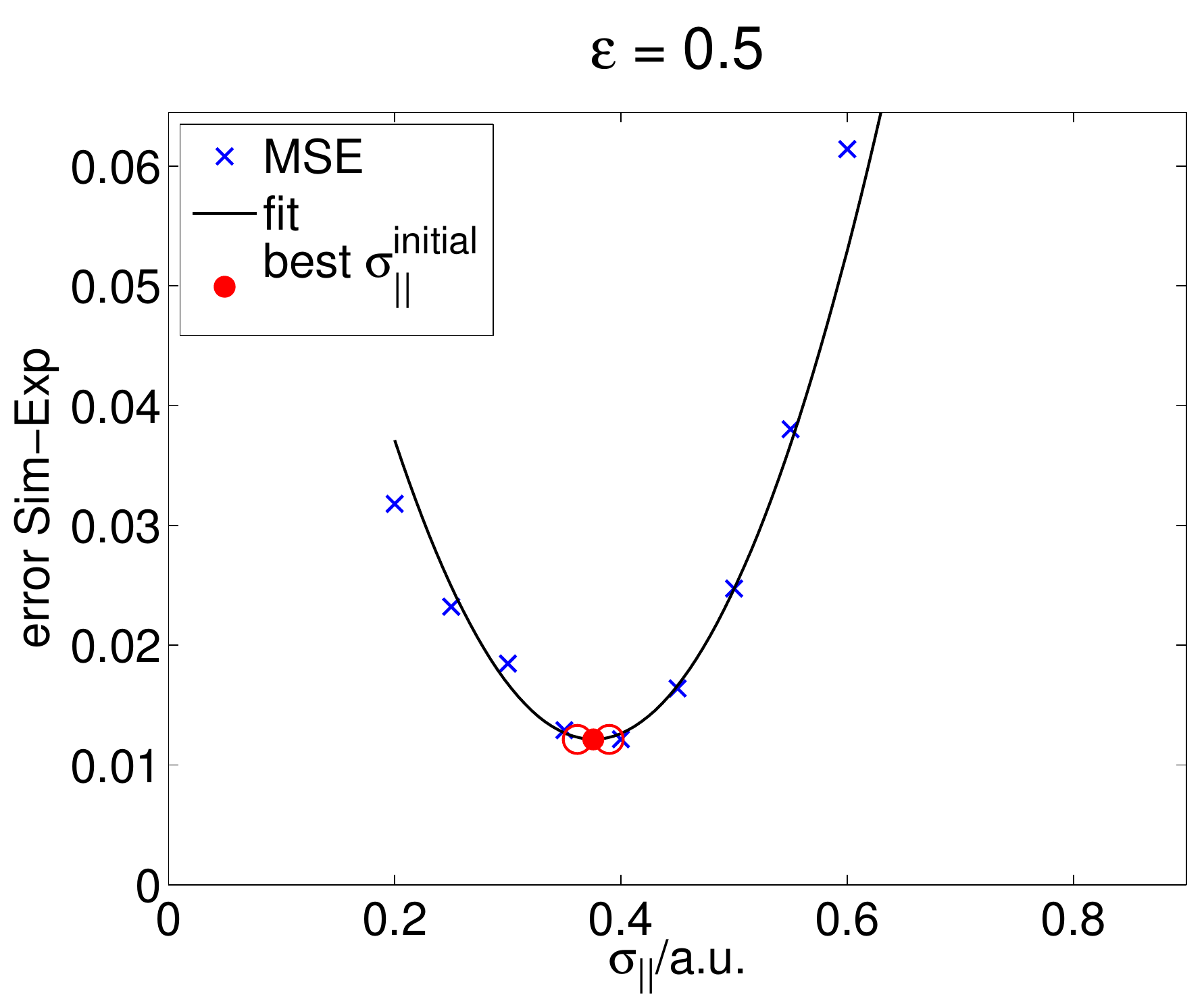}
\caption{\textbf{Simulation error depending on initial longitudinal momentum spread.} Evolution of the mean square errors (blue $\times$) between simulations and experimental data for increasing initial longitudinal momentum spreads in the simulations, calculated from elliptical segment integration. The black solid line shows the quadratic curve fitted through the errors, and the red \fullcircle its minimum point. The red \opencircle indicate the confidence interval for the minimum position. }
\label{fig:e50_ErrorInterpEllr}
\end{figure}

\clearpage
Figure \ref{fig:e5_figComp} shows how the initial longitudinal spread is extracted by comparing semiclassical simulations to experimental data.   
For example, an initial longitudinal momentum spread $\sigma_{||}^{\mathrm{initial}} = 0.4 \mathrm{\, a.u.}$ results in good agreement between the simulation and the experiment. Both lower and higher spreads (exemplified with $\sigma_{||}^{\mathrm{initial}} = 0 \mathrm{\, a.u.} \text{ and } 0.8 \mathrm{\, a.u.}$) result in less agreement.  The mean square error for each simulation with the corresponding experimental data was computed. A quadratic curve was fitted through the errors to find the best fitting initial longitudinal momentum spread. The minimum point of the fitted curve was accepted as the best initial longitudinal momentum spread to reproduce the experiment. Figure \ref{fig:e50_ErrorInterpEllr} shows the calculated simulation errors for $\epsilon = 0.5$ with the elliptical integration. 
The confidence interval for the minimum parameter of the quadratic fit to the set of errors was taken as a lower limit error bar length for the respective technique.

\subsection{Results}
\begin{figure}[h!]
\centering
\includegraphics[width=\textwidth]{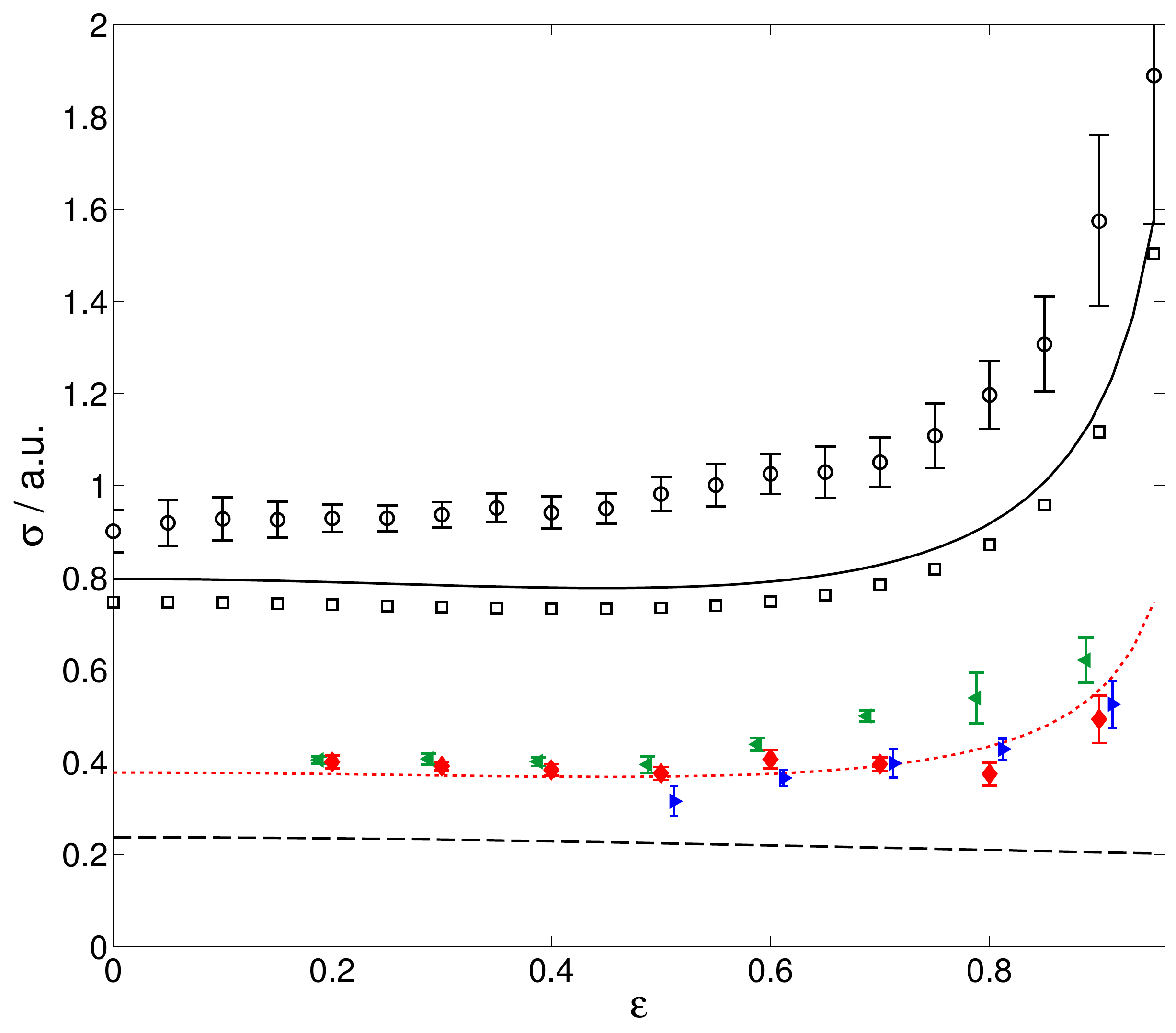}
\caption{\textbf{Comparison of initial and final longitudinal momentum spreads.} 
Black $\opencircle$ are the longitudinal momentum spread extracted from the experimental data using elliptical integration. The black solid line is the analytical formula \eref{eq:sigmaLTheo} for the final longitudinal momentum spread. The black $\opensquare$ show the analytical longitudinal spread averaged over the decreasing intensity for all optical cycles in the pulse \eref{eq:sigmaLTheoAverage}.
The best fitting values from the simulation are marked in blue $\blacktriangleright$ slightly shifted to the right using the radial integration for $\epsilon > 0.4$, green $\blacktriangleleft$ slightly shifted to the left using the $x$ projection, and red $\fulldiamond$ using the elliptical segments technique. All of the error bars show the confidence interval for 98\% of the corresponding fit parameter. The values found show a similar trend as the analytical formula \eref{eq:sigmaLTheo} indicated by the red dotted line, shifted down to fit the red values. The black dashed line at the bottom is the analytical formula \eref{eq:vtransProb} for the transverse momentum spread.}
\label{fig:sigmaLBest}
\end{figure}

The longitudinal momentum spread at the tunnel exit, calculated using three different methods (radial integration, projection and elliptical integration), are shown in figure \ref{fig:sigmaLBest}, along with the theoretically calculated longitudinal spread in \eref{eq:sigmaLTheo}, and the experimental longitudinal spread recorded at the detector. The error bars show confidence intervals for a confidence level of 0.98 for the extraction of the minimum point in the quadratic fit to the simulation-experiment errors. 

For values of $\epsilon$ between 0.2 and 0.9, the analysis was performed. As mentioned before, only when the polarization $|\epsilon|<1$ breaks the rotation symmetry, the longitudinal spread can be studied. For $\epsilon < 0.2$, there are too many electrons which come closer than 5 a.u. to the parent ion and have to be discarded in the simulation, such that a direct comparison of the calculated distribution to the measured data is not reasonable. 

The general trend of the new simulations shows a different behaviour to the one given in \cite{Wavepacket}. The values for $\sigma_{||}^{\mathrm{initial}}$ lie almost on a curve given by \eref{eq:sigmaLTheo} shifted down to an appropriate level.

\subsection{Effective ionization potential}
To test whether the additional longitudinal spread is caused by different exit points, we use an effective ionization potential given by \cite{Murray2010}
\begin{equation}
I_{\mathrm{p,eff}} = I_{\mathrm{p}} + \frac{(v_{\bot}^{\mathrm{initial}})^2}{2} \label{eq:Ieff}.
\end{equation}
Using this effective ionization potential in the calculation of the tunnel exit makes the exit point dependant on the initial transverse momentum of the respective electron at the tunnel exit, introducing an additional spread in the final longitudinal momenta distribution.  To study the significance of this effect, the initial longitudinal momentum spread was set to $0 \mathrm{\, a.u.}$ for all simulation runs, and the exit point for each electron defined using the above equation.  It was found however, that including different exit radii in the simulations introduced only a small additional spread that does not account for the experimentally measured spread found in \cite{Wavepacket}. 

\begin{figure}[h]
\centering
\includegraphics[width=\textwidth]{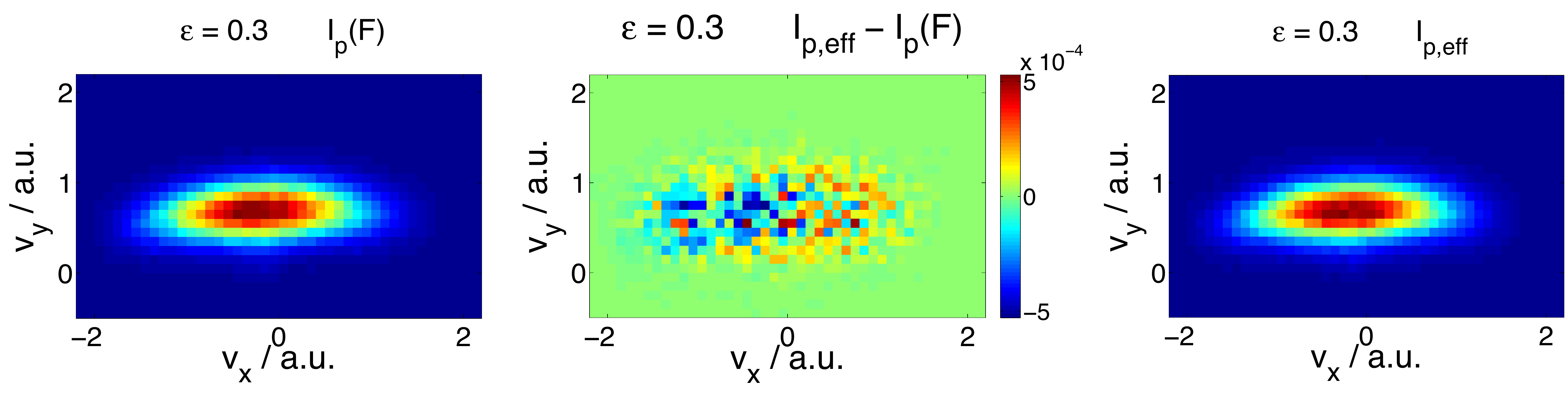}
\caption{\textbf{Effective ionization potential.} Calculating the difference (middle) of the momentum distribution with the effective ionization potential \eref{eq:Ieff} (right) and with the Stark shifted ionization potential (left) reveals a shift of the momentum distribution in positive $v_x$ direction for ellipticity $\epsilon = 0.3$.}
\label{fig:IpEff_03}
\end{figure}

 A comparison of the results with and without effective ionization potential also shows a slight shift in the momentum distribution towards higher $v_x$, see for example figure \ref{fig:IpEff_03} for the case of $\epsilon = 0.3$. 

\begin{figure}[ht]
\centering
\includegraphics[width=\textwidth]{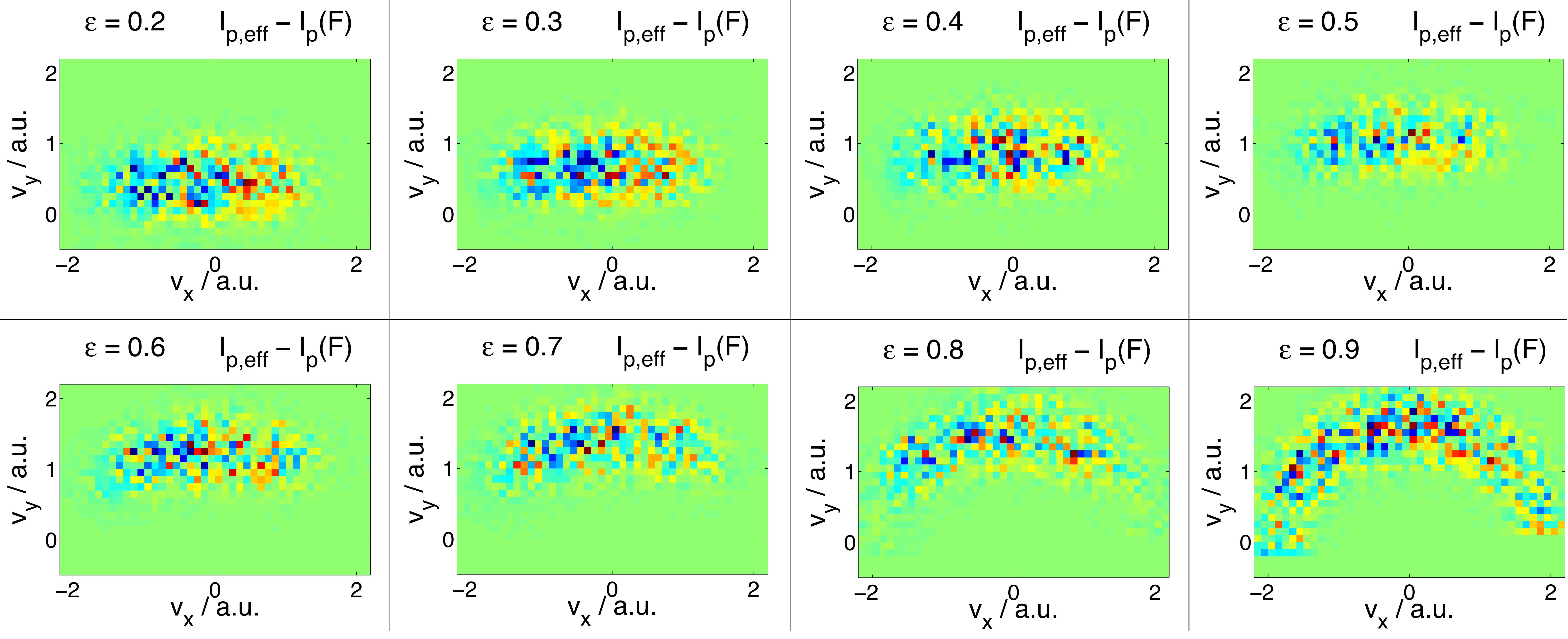}
\caption{\textbf{Effective ionization potential, ellipticity scan.} For lower ellipticity, the difference in distributions reveals a shift of the momentum distribution in positive $v_x$ direction. With increasing ellipticity however, the momentum distributions with or without effective ionization potential overlap more, until only the increased transverse momentum spread is visible. Yellow or red colours indicate higher probability from $I_{\mathrm{p,eff}}$, blue tones indicate higher probability from $I_{\mathrm{p}}(F)$.}
\label{fig:IpEff}
\end{figure}

But the shift vanishes for $\epsilon > 0.5$, as demonstrated in the evolution of the differences in figure \ref{fig:IpEff}.

\section{Discussion}
\subsection{Double peak structure in angular distribution}
\begin{figure}[htb]
\centering
\includegraphics[width=\textwidth]{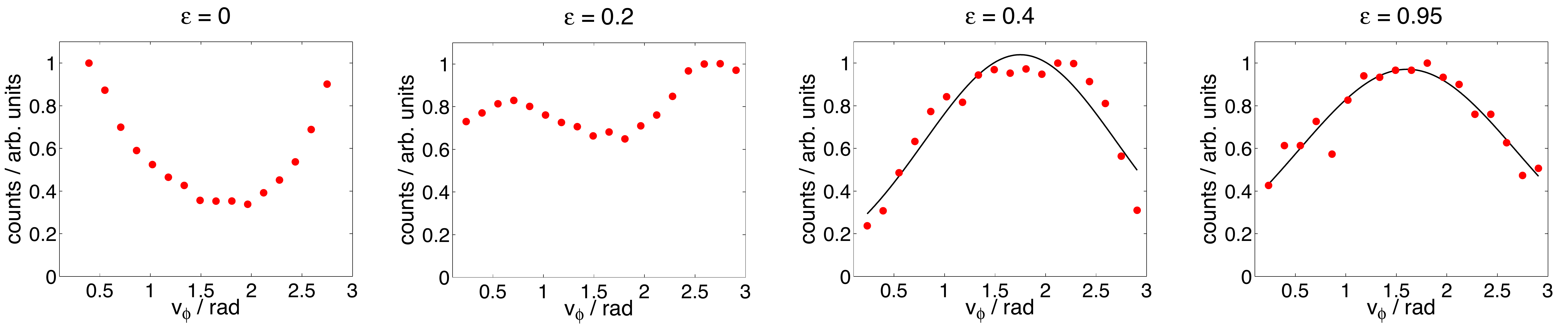}
\caption{\textbf{Angular momentum distribution.} The angular distributions arising from radial integration (red points) of the top half distributions shown above in figure \ref{fig:MomScanEllipticity} exhibit a double peak structure for small ellipticity $\epsilon$. Only for $\epsilon$ large enough can a Gaussian fit (black solid line) reproduce the distribution. The angle $\phi$ is counted from the $x$-axis anticlockwise.}
\label{fig:figCompP}
\end{figure}

When analysing the velocity distribution observed at the detector in \cite{Wavepacket} using a radial integration, a double peak structure was found in the angular distribution for $\epsilon \leq 0.35$. Only for $\epsilon>0.4$, does the angular distribution approach a Gaussian, see figure \ref{fig:figCompP}.
This double peak structure is a result of radial integration and can be understood as follows.

In figure \ref{fig:MomScanEllipticity} for the case of $\epsilon=0.2$, black rays from the origin of the coordinate system are plotted for two example angles. It can be seen that depending on the angle, a different number of pixels with significant count rate are integrated over. Additionally, for angles lying close to the $x$-axis, the integration goes over longitudinal momentum as well. This effect is of course stronger for smaller ellipticities, because the centres of distribution are then closer to the origin of the coordinate system, and consequently the difference in inclusion of longitudinal momentum is more pronounced. This explains why the double peak structure was only found for small enough ellipticity in \cite{Wavepacket}.  In other words, the angular momentum distribution calculated from radial integration finds a spurious double peak which is due to the elliptic geometry of the momentum distribution, and it only appears for ellipticity in a range where radial integration strongly mixes the transverse and the longitudinal components of the momentum distribution.

\begin{figure}[htb]
\centering
\includegraphics[width=\textwidth]{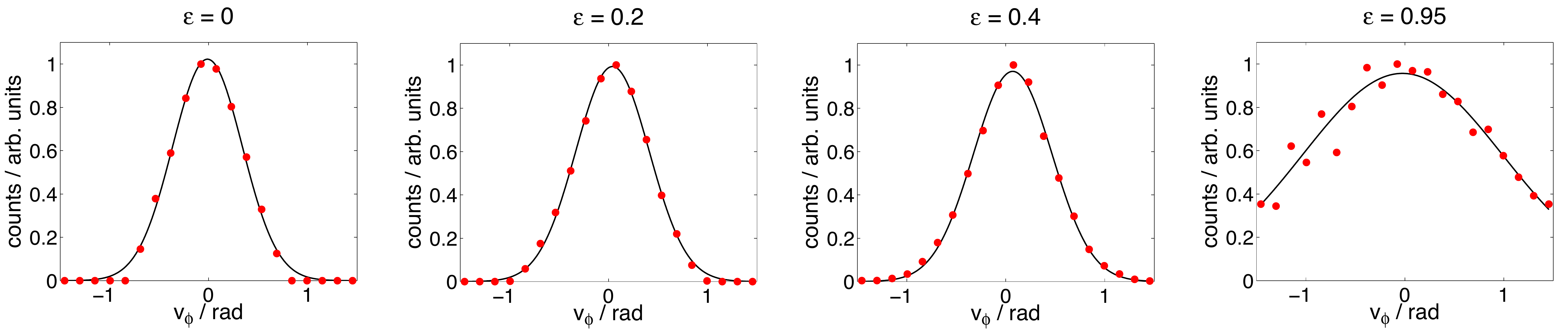}
\caption{\textbf{Momentum distribution obtained using elliptical integration method.} Using elliptical segment integration (red points) on the top half momentum distributions shown above in figure \ref{fig:MomScanEllipticity} yields angular distributions which are well-reproduced by Gaussian fits (black solid line) for any ellipticity $\epsilon$. The angle $\phi$ is counted from the $y$-axis anticlockwise.}
\label{fig:figCompE}
\end{figure}

Taking into account this elliptical geometry of the momentum distribution using the above defined elliptical segments results in a well-defined Gaussian distribution for any ellipticity, as demonstrated in figure \ref{fig:figCompE}.

\subsection{Initial longitudinal momentum spread}
The plotted error bars in figure \ref{fig:sigmaLBest} show the confidence interval from the fitted curve through the simulation errors depending on the initial longitudinal momentum spread, with a confidence level of 98\%. For ellipticity too small, the estimated errors from the radial integration were sometimes so large that they would have covered more than the range of the plot. The $x$ projection seemed to be more stable even for large ellipticities, the error bars did grow considerably, but remained below $\pm 0.06 \mathrm{\, a.u.}$ at maximum. The elliptical segment integration however produces accurate results for all ellipticities. For this reason, also the final longitudinal momentum spread from the experimental data was extracted using this technique. 

The analytical formula \eref{eq:sigmaLTheo} does not take into account that the peak field strength for optical cycles which are not at the centre of the pulse is weaker than the overall peak strength.  Averaging over weaker field strengths in both experiment and simulation leads to a smaller longitudinal spread acquired during propagation in the laser field. This in turn makes the difference between the theoretical prediction in \eref{eq:sigmaLTheoAverage} and the experiment in \cite{Wavepacket}
even larger.

All means of data analysis presented here (radial integration, projection and elliptical integration) agree that the fit between the experiment and simulation is best if an initial longitudinal momentum spread is included. The values of $\sigma_{||}^{\mathrm{initial}}$ which yield the best fit are roughly around 0.4 a.u., and therefore considerably bigger than the detector resolution.  

Furthermore, the values of initial longitudinal momentum spread show a strikingly similar behaviour to the analytical formula \eref{eq:sigmaLTheo}. Fitting that function through the values obtained by elliptical integration results in a reasonable agreement. However, to the best of our knowledge there is no physical model that would justify this agreement. 

\begin{figure}[htbp]
\centering
\includegraphics[width=0.35\textwidth]{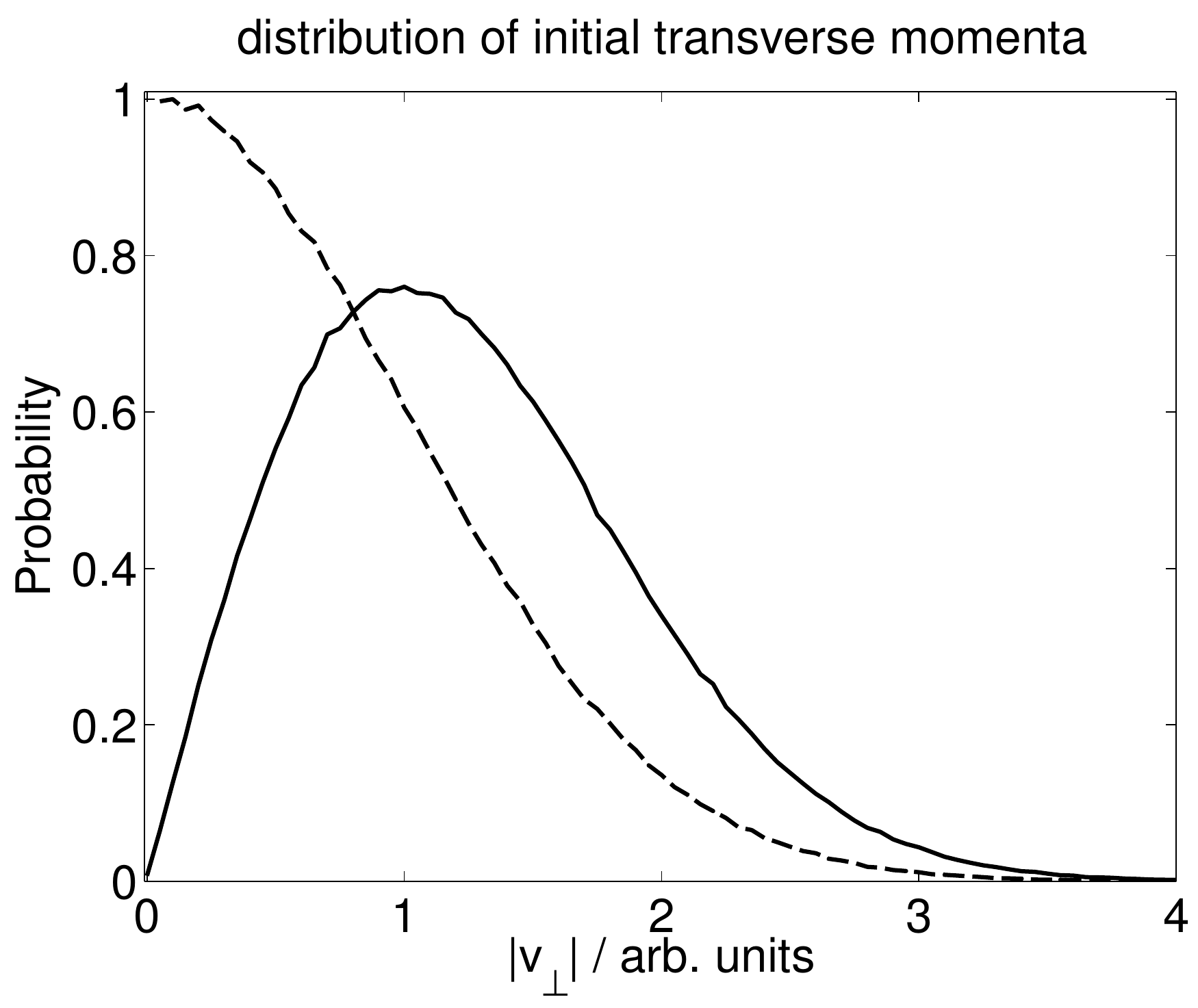}
\caption{\textbf{Initial transverse momentum spread.} The absolute value of the initial transverse momentum used in \cite{Wavepacket} given by \eref{eq:narrowAbsDistro} (dashed line) and in this new simulation obtained from \eref{eq:TransMomComp} using \eref{eq:ADKtransverse} (solid line) show different distributions. The average transverse momentum in the new simulation is higher.}
\label{fig:distAbs}
\end{figure}

The $\sigma_{||}^{\mathrm{initial}}$ values which have been previously published in \cite{Wavepacket} are somewhat higher than the values found from this simulation.
For small ellipticity, this can be attributed to the radial integration technique, which was used exclusively to analyse the results in \cite{Wavepacket}. 
Additionally, the initial transverse momentum spread was generated as described in \cite{Liu2010}, which assumes a transverse momentum probability
\begin{equation}
P(|v_{\bot}^{\mathrm{initial}}|) \propto \exp\left(-\frac{(v_{\bot}^{\mathrm{initial}})^2}{2\sigma_{\bot}^2}\right). \label{eq:narrowAbsDistro}
\end{equation}
The spread resulting from this distribution is narrower than the one from \eref{eq:TransMomComp}, figure \ref{fig:distAbs} illustrates the difference.
The discrepancy is due to the fact that the probability distribution must be transformed as an integral over all $v_{\bot,\mathrm{ip}},v_{\bot,\mathrm{op}}$ which fulfil $v_{\bot}^2 = (v_{\bot,\mathrm{ip}}^{\mathrm{initial}})^2 + (v_{\bot,\mathrm{op}}^{\mathrm{initial}})^2$, yielding
\begin{equation}
P(|v_{\bot}^{\mathrm{initial}}|) \propto\exp\left(-\frac{(v_{\bot}^{\mathrm{initial}})^2}{2\sigma_{\bot}^2}\right) 2\pi v_{\bot}. \label{eq:DistroAbsIntegrated}
\end{equation}
The above distribution is valid if 2D CTMC simulation is employed, where the two dimensions correspond to the coordinate along the major axis of polarization and the perpendicular radial coordinate.

\section{Conclusion}
We present an analysis of the longitudinal momentum distribution using a method of elliptical integration.  Unlike radial integration (projection) methods, which mix the longitudinal and transverse components of the momentum spread at low (high) $\epsilon$, this method is robust over the complete range of ellipticity.  With the elliptical integration method, the longitudinal momentum spreads are well-reproduced by a Gaussian over the entire ellipticity range, avoiding the double-peak structure that is observed when using radial integration at low $\epsilon$. 
In agreement with \cite{Wavepacket}, we find that including an initial longitudinal momentum spread at the tunnel exit accounts for our experimental results. This is in contradiction to standard theoretical assumptions.  Further theoretical work is necessary to find a physical model that accounts for these results.

\ack
This work was supported by NCCR Quantum Photonics (NCCR QP) and NCCR Molecular Ultrafast Science and Technology (NCCR MUST), research instruments of the Swiss National Science Foundation (SNSF), by ETH Research Grant No. ETH-03 09-2, an SNSF equipment grant and a Marie Curie International Incoming Fellowship within the 7th European Community Framework Programme (grant no. 275313). Our ultrafast activities are supported by the ETH Femtosecond and Attosecond Science and Technology (ETH-FAST) initiative as part of the NCCR MUST program.

\section*{References}
\bibliographystyle{unsrt} 
\bibliography{JoPB_References}

\vspace{1cm}
\noindent
{\footnotesize published in \\J. Phys. B: At. Mol. Opt. Phys. 46 (2013) 125601 \href{http://dx.doi.org/10.1088/0953-4075/46/12/125601}{doi:10.1088/0953-4075/46/12/125601}}

\end{document}